\documentclass[twocolumn,preprintnumbers,amsmath,amssymb]{revtex4}
\usepackage{graphicx}
\usepackage{amssymb}

\usepackage{multirow}

\begin{document}

\title{Anomalous nonlocality of information masked in quantum correlations}
\author{Guang Ping He}
\email{hegp@mail.sysu.edu.cn}
\affiliation{School of Physics, Sun Yat-sen University, Guangzhou 510275, China}

\begin{abstract}
Although information, strictly speaking, is not a physical entity, it generally requires physical entities as its carriers, e.g., writing it down on paper, encoding
it with quantum particles, or transmitting it using electro-magnetic fields.
And it seems natural that these carriers cannot travel faster than light.
Here we reveal that if we use quantum correlations as the carrier of
information (either quantum or classical), then it can display a kind of
nonlocality, which bears both similarities to and distinctions from the
nonlocality of physical particles. Notably, though superluminal signaling is
still not allowed so that the special relativity is not violated, it is
possible to select at our will whether to decode the information at one
location, or to dispatch it to another location far away (i.e., to give up
the chance of decoding the information and let it be decodable in somewhere
else only) without needing the assistance of classical information, so that
it occurs instantaneously without being limited by the speed of light. This
phenomenon differs sharply from the nonlocality of physical
particles that we once knew, where whether a particle can be detected in one
location or another is governed by quantum uncertainty, which cannot be
chosen freely.
\end{abstract}

\maketitle



Nonlocality is no stranger to quantum mechanics. It allows events
separated by space to influence each other instantaneously. For example, a
quantum particle can have nonvanishing probabilities to be detected at
different locations in space. But once it was detected at one location, the
probabilities for finding it at other locations drop to zero immediately. A
distinct feature in this example is that the observer cannot determine at
his choice whether the particle can be found at a given location or not.
Instead, it is governed by quantum uncertainty. On the other hand,
information is generally acompanied with mass and/or energy. It is therefore
natural to expect that information displays behaviors consistent with those
of the physical carriers. In this study, we show that by using quantum
correlations as carriers, information can also exhibit a form of
nonlocality. This is achieved by leveraging a recently proposed encrypted
cloning scheme adapted for information masking. But more intriguingly, while
this nonlocality has many similarity with that of particles, there is a
notable difference. That is, the observer can choose freely whether to
decode the information locally, or dispatch it instantaneously to another
arbitrarily distant spatial location, though it does not enable superluminal
signal transfer. In this context, \textquotedblleft
dispatch\textquotedblright\ means waiving the possibility of decoding the
information at the original site and rendering it decodable exclusively at
the distant position. Our result reveals the diversity of nonlocality, and
the profounding relationship between information and physical substances.

\section*{The masking scheme}

Masking information \cite{qbc140,qi1723,qi1895} means encoding information
in composite quantum states, so that it is completely hidden from the
reduced subsystems and can be found only in the correlations. It was proven
that quantum information cannot be masked in two-partite correlation, but
can be masked in multi-partite correlation \cite{qbc205,qi1840,qi2446}.

Very recently, an encrypted cloning scheme was proposed \cite{qi2564} and
implemented experimentally \cite{qi2594}. Here we put aside its original
purpose for encrypted cloning, and focus on the fascinating properties of
the scheme when adapted for information masking. Using the $n=2$ case of the
encrypted cloning scheme in \cite{qi2564} as an example, information can be
masked in three-partite quantum correlation as follows:

Consider that we have two pairs of qubits $S_{j}$ and $N_{j}$ ($j=1,2$)
initialized in the Bell state%
\begin{equation}
\left\vert \phi \right\rangle _{S_{j}N_{j}}=\frac{1}{\sqrt{2}}(\left\vert
0\right\rangle _{S_{j}}\left\vert 0\right\rangle _{N_{j}}+\left\vert
1\right\rangle _{S_{j}}\left\vert 1\right\rangle _{N_{j}}),  \label{Bell}
\end{equation}%
and a qubit $A$ carrying the information (either known or unknown) to be
masked, which can be written as the quantum state%
\begin{equation}
\left\vert \psi \right\rangle _{A}=a\left\vert 0\right\rangle
_{A}+b\left\vert 1\right\rangle _{A}.  \label{original}
\end{equation}%
Note that this information is not limited to quantum ones. If we want to
mask a classical bit $0$ or $1$, we can simply initialize the state of qubit
$A$ in $\left\vert 0\right\rangle _{A}$ or $\left\vert 1\right\rangle _{A}$,
respectively.

Following Eqs. (6) and (7) of \cite{qi2564}, the encoding operation on the
combined system $A\otimes (S_{1}\otimes N_{1})\otimes (S_{2}\otimes N_{2})$\
takes the form%
\begin{equation}
U_{enc}=\frac{1}{2}\sum\limits_{\mu =0}^{3}\alpha _{\mu }^{-1}\sigma _{\mu
}^{\left( A\right) }\otimes \left( \sigma _{\mu }^{\left( S_{1}\right)
}\otimes I^{\left( N_{1}\right) }\right) \otimes \left( \sigma _{\mu
}^{\left( S_{2}\right) }\otimes I^{\left( N_{2}\right) }\right) ,
\end{equation}%
and the postencoding state becomes%
\begin{eqnarray}
&&\left\vert \Gamma \right\rangle _{A\otimes (S_{1}\otimes N_{1})\otimes
(S_{2}\otimes N_{2})}  \nonumber \\
&=&U_{enc}\left[ \left\vert \psi \right\rangle _{A}\otimes \left\vert \phi
\right\rangle _{S_{1}N_{1}}\otimes \left\vert \phi \right\rangle
_{S_{2}N_{2}}\right]   \nonumber \\
&=&\frac{1}{2}\sum\limits_{\mu =0}^{3}\alpha _{\mu }^{-1}\sigma _{\mu
}^{\left( A\right) }\left\vert \psi _{A}\right\rangle \otimes \left\vert
\phi _{\mu }\right\rangle _{S_{1}N_{1}}\otimes \left\vert \phi _{\mu
}\right\rangle _{S_{2}N_{2}},  \label{post}
\end{eqnarray}%
where $I^{(k)}$ and $\sigma _{\mu }^{(k)}$ ($\mu =0,1,2,3$) are the identity
operator and the Pauli operators, respectively, acting on the system $k$\ ($%
k\in \{A,S_{1},N_{1},S_{2},N_{2}\}$), $\alpha _{0}=1$, $\alpha _{1}=\alpha
_{2}=\alpha _{3}=i$, and%
\begin{equation}
\left\vert \phi _{\mu }\right\rangle _{S_{j}N_{j}}=\sigma _{\mu
}^{(S_{j})}\otimes I^{(N_{j})}\left\vert \phi \right\rangle _{S_{j}N_{j}}.
\end{equation}

\section*{Properties}

Now we group the postencoding state into three systems:

System $X$: the postencoding qubit $A$;

System $Y$: qubits $S_{1}$ and $N_{1}$;

System $Z$: qubits $S_{2}$ and $N_{2}$.

These systems have the following notable properties:

(i) Any single one of the three systems does not contain the masked
information $\left\vert \psi \right\rangle _{A}$ of the original qubit $A$
(i.e., Eq. (\ref{original})). This can be proven by calculating the reduced
density matrix of each system via Eq. (\ref{post}), as shown in \cite{qi2564}%
.

(ii) Any two of the three systems is sufficient to recover the masked
information. For example, if we have access to systems $Y$ and $Z$, then
following Eq. (5) of \cite{qi2564}, the masked information of the original
qubit $A$ can be recovered on qubit $S_{1}$ by applying on $S_{1}\otimes
N_{1}\otimes S_{2}\otimes N_{2}$ with the decoding operation%
\begin{equation}
U_{dec}^{(YZ)}=\sum\limits_{\mu =0}^{3}\alpha _{\mu }\left( \left\vert \phi
_{\mu }\right\rangle \left\langle \phi _{\mu }\right\vert
_{S_{1}N_{1}}\right) \otimes I_{S_{2}}\otimes \sigma _{\mu }^{\left(
N_{2}\right) \intercal }
\end{equation}%
where $\sigma _{\mu }^{\left( N_{2}\right) \intercal }$\ denotes the
transpose of $\sigma _{\mu }^{\left( N_{2}\right) }$. Or if we have access
to systems $X$ and $Y$ (or $X$ and $Z$), then the masked information can be
recovered on qubit $A$ by applying on $A\otimes S_{1}\otimes N_{1}$ (or $%
A\otimes S_{2}\otimes N_{2}$) with the decoding operation

\begin{equation}
U_{dec}^{(XY)}=\sum\limits_{\mu =0}^{3}\alpha _{\mu }\sigma _{\mu }^{\left(
A\right) }\otimes \left( \left\vert \phi _{\mu }\right\rangle \left\langle
\phi _{\mu }\right\vert _{S_{1}N_{1}}\right)  \label{UXY}
\end{equation}%
or

\begin{equation}
U_{dec}^{(XZ)}=\sum\limits_{\mu =0}^{3}\alpha _{\mu }\sigma _{\mu }^{\left(
A\right) }\otimes \left( \left\vert \phi _{\mu }\right\rangle \left\langle
\phi _{\mu }\right\vert _{S_{2}N_{2}}\right) ,
\end{equation}%
respectively.

(iii) Once the masked information is decoded from two of the three systems,
decoding it from anywhere else becomes impossible. That is, only one copy of
the information can be decoded, which is in alignment with the quantum
no-cloning theorem \cite{qi15}.

\section*{Nonlocality characters}

Now we focus on the nonlocality characters of the information masked in the
scheme.

The above property (i) means that the information $\left\vert \psi
\right\rangle _{A}$ of the original qubit $A$ is not stored in any of the
three systems. Therefore, it is stored only in the correlations between
these systems. For concreteness, in the following we call these correlations
as correlation $XY$, correlation $XZ$, and correlation $YZ$, respectively,
as labelled in Fig.1(a). Then properties (ii) and (iii) reveal that the
information $\left\vert \psi \right\rangle _{A}$ displays the following
nonlocality:

\begin{figure*}[thbp]
\includegraphics[scale=0.7]{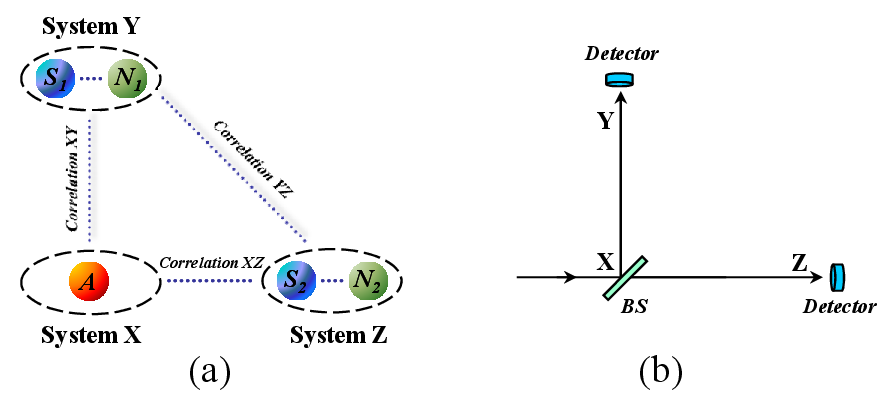}
\caption{Comparison between the nonlocality of masked information and
the nonlocality of physical particles. (a) The masked information is encoded
in correlation $XY$, correlation $XZ$, and correlation $YZ$. We can either
decode the information from the correlation between any two systems (e.g.,
correlation $XY$), or dispatch it to somewhere else in space (e.g.,
correlation $YZ$) at our choice. (b) A single photon impinges on a 50/50
non-polarizing beam splitter (BS) at point $X$. Two detectors are placed at
the end of the reflection path $XY$ and the transmission path $XZ$,
respectively. The photon can be detected either at path $XY$ or at path $XZ$
with equal probabilities. But we cannot choose deterministically at which
path it will be detected.}
\label{fig:epsart}
\end{figure*}

(I) Before applying any decoding operation, any of the correlations $XY$,
correlation $XZ$, and correlation $YZ$ contains the information.

(II) Once the information is decoded from any of the correlations, e.g.,
correlation $XY$, then it immediately disappears from correlations $XZ$ and $%
YZ$. This phenomenon occurs no matter how far system $Z$ is separated from
systems $X$ and $Y$ in space, and does not take any classical signal
transmission to complete, so that it is not limited by the speed of light,
just as the \textquotedblleft spooky action at a distance\textquotedblright\
named by Einstein.

(III) We can select whether to decode the information at one location or
dispatch it to other locations \textit{at our own choice}. That is, suppose
that we own both systems $X$ and $Y$. Then we can apply the operation $%
U_{dec}^{(XY)}$\ in Eq. (\ref{UXY}) to decode the information. On the other
hand, if we own system $X$ and we destroy the quantum correlation between it
and other systems (e.g., we perform a local measurement on it in any basis,
and discard the measurement outcome without keeping a record or announcing
it classically), then we give up the chance of decoding the masked
information, and this information is completely transferred to correlation $%
YZ$ instantaneously, no matter how far away the systems $Y$ and $Z$ are.

To sum up, before decoding, the three scenarios \textquotedblleft the
information is stored in correlations $XY$\textquotedblright ,
\textquotedblleft the information is stored in correlations $XZ$%
\textquotedblright , and \textquotedblleft the information is stored in
correlations $YZ$\textquotedblright\ exist simultaneously, similar to the
superposition of quantum states. Once the information is decoded from one of
the correlations, then the other scenarios immediately vanish, which is also
similar to the collapse of quantum superposition. However, the most
important character we found here, is that we have the freedom to choose
which of the three scenarios will \textquotedblleft
collapse\textquotedblright\ to, at our own choice! This is in sharp contrast
to the collapse of quantum superposition, where the outcome is fully
governed by quantum uncertainty, leaving no room for arbitrary choice of the
outcome.

\section*{Differences from the nonlocality of physical particles}

To gain a deeper understanding on the significance of the above characters,
especially character (III), now let us further compare the nonlocality of
masked information with the nonlocality in other quantum phenomena that we
are already familiar with.

Fig.1(b) shows a typical setup in quantum optics. A single photon impinges
on a 50/50 non-polarizing beam splitter at point $X$. Two detectors are
placed at the end of the reflection path $XY$ and the transmission path $XZ$%
, respectively. {}It is a well-known result that if the photon is detected
at path $XY$, then we can immediately infer that it does not exist at path $%
XZ$.{} Or if the detection at path $XY$ returns null result, then we know
that the photon must exist at path $XZ$. But we cannot control with
certainty whether the photon will be found at path $XY$ or path $XZ$. On the
contrary, in the nonlocality of the masked information as ilustrated in
Fig.1(a), we can always determine whether to recover the information from
correlations $XY$, or to dispatch it to correlations $YZ$, without being
constrained by quantum uncertainty anymore.

It is also worth noting that quantum teleportation \cite{qi179} is another
approach for sending quantum information via correlation, but it also
exhibits a sharp difference compared to the case we described. In brief, the
process of quantum teleportation goes as follows. Suppose that Alice has a
qubit $A$ in the state $\left\vert \psi \right\rangle _{A}$, and they share
a Bell state in the form of Eq. (\ref{Bell}) where Alice has qubit $S$ and
Bob holds qubit $N$. To teleport the state $\left\vert \psi \right\rangle
_{A}$ to Bob, the Alice measures qubits $A$ and $S$ collectively in the Bell
basis, and announces the measurement result to Bob classically. Based on the
announced classical information, Bob performs a corresponding unitary
operation on qubit $N$ so that its state turns into $\left\vert \psi
\right\rangle _{A}$. That is, Alice can also choose to decode the
information $\left\vert \psi \right\rangle _{A}$ of the original qubit $A$
herself (by keeping qubit $A$ unmeasured) or dispatch this information to
Bob (by measuring $A$ and $S$ collectively in the Bell basis). But an
important component of the latter process is: classical communication is
necessary to complete the transmission of the quantum information to Bob. If
Alice does not announce her measurement result classically, Bob does not
receive the information $\left\vert \psi \right\rangle _{A}$ even though
this information no longer exists at Alice's side after her performed the
Bell measurement. This is because, at this stage, the reduced density matrix
of Bob's qubit $N$, tracing over all possible results of Alice's
measurement, contains no information about Alice's $\left\vert \psi
\right\rangle _{A}$. Thus, the speed of completing the transmission of
quantum information in teleportation is limited by the speed of classical
communication. But in our case, if Alice holds system $X$ and she wants to
dispatch the masked information to correlation $YZ$, all she needs is to
measure system $X$ locally in an arbitrary basis so that the correlation
between $X$ and other systems is destroyed. No classical communication is
required and the masked information \textquotedblleft
collapses\textquotedblright\ to correlation $YZ$ instantaneously,
unrestricted by the speed of classical communication. Nevertheless, we
should also note that the transferred quantum information $\left\vert \psi
\right\rangle _{A}$ is masked in the correlations $XY$, $XZ$, and $YZ$
beforehand, which cannot be altered at the moment when Alice decided to
dispatch it to correlation $YZ$. Therefore, it will not enable superluminal
signal transfer, so that it does not conflict with special relativity.

Finally, as addressed in \cite{qi2564} and \cite{qi1942}, respectively, both
the encrypted cloning and information masking schemes have impacts on the
theory of black hole. Our result is no exception. Consider the information
masked in the correlations in the above three systems. When one of the
system is falling into a black hole so that any information stored in this
system is completely eliminated, the masked information will escape to the
correlation between the rest two systems instaneously, making the
information survive automatically. This phenomenon may somewhat deepen our
understanding on the philosophy why masking quantum information in
two-partite correlation using unitary transformation is impossible \cite%
{qbc140}. That is, similar to the vibe of quantum Darwinism \cite{qi822},
information masked in multi-partite correlation instead of two-partite
correlation can increase the chance of survival.

\section*{Acknowledgements}

This work was supported in part by Guangdong Basic and Applied Basic
Research Foundation (Grant No. 2019A1515011048).

\end{document}